\documentstyle[12pt]{article}
\begin{document}
\setlength{\baselineskip}{0.30in}
\newcommand{\beq}{\begin{equation}}
\newcommand{\eeq}{\end{equation}}
\newcommand{\bi}{\bibitem}
\def\mpl{m_{Pl}}

\begin{flushright}
UM - TH - 94 - 14\\
May 13, 1994\\
gr-qc/9405026
\end{flushright}

\begin{center}
\vglue .06in
{\Large \bf {The Vacuum of de~Sitter Space}}\\[.5in]

{\bf A. D. Dolgov,\footnote{Permanent address: ITEP, 113259, Moscow,
Russia.} M. B. Einhorn, and V. I. Zakharov}
\\[.05in]
{\it{The Randall Laboratory of Physics\\
University of Michigan\\
Ann Arbor, MI 48109}}\\[.15in]
\end{center}
\begin{abstract}
\begin{quotation}
To resolve infrared problems with the de~Sitter invariant vacuum, we argue that
the history of the de~Sitter phase is crucial.  We illustrate how either
(1)~the diagonalization of the Hamiltonian for long-wavelength modes or (2)~an
explicit modification of the metric in the distant past leads to natural
infrared cutoffs.  The former case resembles a bosonic superconductor in which
graviton-pairing occurs between non-adiabatic modes.  While the dynamical
equations respect de~Sitter symmetry, the vacuum is not de~Sitter invariant
because of the introduction of an initial condition at a finite time.  The
implications for the one-loop stress tensor and the production of particles are
also discussed.

\end{quotation}
\end{abstract}
\newpage
\section{Introduction}

One of the most perplexing problems of cosmology concerns the magnitude of the
cosmological constant,\footnote{For a review, see e.g. Ref.~\cite{sw,ad}.} for
which a stringent observational upper limit exists.
Although supersymmetry requires the cosmological constant to vanish,
supersymmetry is either explicitly or spontaneously broken in nature.
\footnote{Spontaneously broken global SUSY demands a nonzero vacuum energy in
the broken phase. In supergravity a nonvanishing $\rho_{vac}$ is not obligatory
and one can adjust the parameters so that it vanishes when the symmetry is
broken but the required fine-tuning is unnatural and should be accurate with
the precision of about $10^{-100}$.} As a result, the low-energy effective
field theory is not supersymmetric, and the natural scale of the cosmological
term in the gravitational action should be no smaller than the magnitude of the
vacuum energy associated with the quanta of matter and gravitons themselves.
As the universe cools, such a cosmological term would eventually predominate
over radiation and other forms of matter.  Indeed, the most popular cosmology
today, the inflationary scenario (for a review see the book \cite{al}), assumes
that our universe passed through an era in which the cosmological term
dominated, and it is a total mystery why, when this phase terminated, we should
be left in a universe with a tiny or vanishing vacuum energy.  We do not
advance a solution to the cosmological constant problem here, but address the
simpler but nevertheless vexing question of what is the correct vacuum state
during the inflationary era.  The answer to this question is a prelude to
addressing consistently the issue of whether such a phase is intrinsically
unstable due to gravitational fluctuations, as some have
suggested.\cite{lf,ait,tw1,tw2,tw4} It also is crucial for understanding the
quantum field theory of a free, massless, minimally coupled scalar field in a
de~Sitter background.\cite{lf}

The general form for the action may be written as the sum of gravitational and
matter terms:
\beq
S=S_g+S_m.
\label{action}
\eeq
The gravitational action $S_g$ with a cosmological constant $\Lambda$ is
\beq
S_g={1\over \kappa^2} \int d^4x \sqrt{g} (R - 2\Lambda),
\label{gravaction}
\eeq
with $\kappa^2 \equiv 16\pi/\mpl^2$.  The corresponding field equations are
\beq
R_{\mu\nu} -{1\over 2} g_{\mu\nu}(R-2\Lambda)={\kappa^2\over2}T_{\mu\nu}.
\label{eineq}
\eeq
where $T_{\mu\nu}$ is the stress tensor of matter.
It is presumed that the cosmological term, which may actually arise in part
from vacuum fluctuations of particles, dominates the stress tensor of these
particles, so we may neglect the matter action and $T_{\mu\nu}$ for the moment.
 de~Sitter space is the maximally symmetric solution of the resulting
equations.  In conformal coordinates appropriate to the spatially flat
sections, the metric takes the form
\beq
g_{\mu\nu}= a^2(\tau) \eta_{\mu\nu},
\label{desitter}
\eeq
 where $\eta_{\mu\nu}$ is the usual Minkowski space metric,
$a(\tau)\equiv-1/H\tau$, with $H=\sqrt{\Lambda/3}$.

The physical fluctuations in the metric, that is to say, the gravitons, in the
neglect of self-interactions, obey the same field equation as the free,
massless, minimally coupled scalar field.\cite{lpg}  Although our ultimate
interest is in the purely gravitational case, for simplicity, we will consider
the quantum field theory of a massless, minimally coupled scalar field in the
de~Sitter background.  This field obeys the same equation of motion as certain
modes of the graviton field, including the physical modes, but avoids the
intricacies of tensor algebra, gauge fixing, unphysical modes and ghosts that
complicate the treatment of the graviton field itself.  In this case, it is
well-known that a de~Sitter invariant vacuum state does not exist.\cite{fp,lf}
To establish notation, we will review the situation briefly.  The action for
the scalar field is simply
\beq
S_m=\int d^4x \sqrt{g} \phi_{,\mu}\phi^{,\mu}\ \ .
\label{matteraction}
\eeq
In the conformal coordinates introduced above, the field equation takes the
form
\beq
\phi^{,\mu}_{,\mu} + {2\partial_\tau a \over a} \partial_\tau\phi =
\phi^{,\mu}_{,\mu}-{2\over\tau} \partial_\tau\phi=0.
\label{eom}
\eeq
The solution of this equation with definite momentum $\vec k$ is
\begin{eqnarray}
\phi_k(\tau,\vec x) =\phi_k(\tau)\exp(i\vec k\cdot\vec
x);& ~~~~ &{\rm with}~\phi_k(\tau)\equiv H\left(\tau-{i\over
k}\right)\exp(-ik\tau )
\label{dswave}
\end{eqnarray}
The corresponding field operator may be written as
\beq
\phi(x)= \int {d^3k \over (2\pi)^3 \sqrt{2k}}\left[
a_{\vec k}^{}\phi_k (\tau)e^{i{\vec k}\cdot{\vec x}} + a_{\vec
k}^{~\dag}\phi(\tau)^*e^{-i{\vec k}\cdot{\vec x}}
\right],
\label{field}
\eeq
with $a_k^{}$ and $a_k^{~\dag}$ satisfying the commutation relations
\beq
\left[a_{\vec k}^{}, a_{\vec q}^{{~\dag}}\right]
=(2\pi)^3 \delta^{(3)} (\vec k -\vec q)
\label{commut}
\eeq
All other commutators are zero.  The natural, de~Sitter invariant no-particle
state $|0,~in\rangle$ that suggests itself is that state annihilated by the
$a_{\vec k}$:
\beq
a_{\vec k}^{}~|0,~in\rangle \equiv 0.
\label{dsvacuum}
\eeq
Unfortunately, although this would appear to be a perfectly acceptable,
adiabatic in-state, this choice of vacuum leads to infrared divergences,
illustrating the difficulties first pointed out in Ref.~\cite{fp}.  For
example, the Feynman propagator in this state is ill-defined because of an
infrared divergence.  The source of the problem is the $i/k$ term in the wave
function, which is a feature not present in Minkowski space (or for the
conformally coupled massless scalar in curved space.)  The propagator takes the
form\cite{tw2}
\beq
G_F(x,x')~=~{H^2\over{4\pi^2}}\left[{ {\tau\tau'}\over{(x-x')^2} }-
{1\over2}\ln(x-x')^2 \right]~.\label{feynman}
\eeq
The scale of the argument of the log is not specified, reflecting the absence
of an infrared cutoff.  This propagator grows for large spacelike separations;
simply inserting a scale into the logarithm would not change that.  In an
acceptable vacuum state, such correlations should fall to zero at large
distances (so-called cluster decomposition.) It is noteworthy that the
commutator of the fields does not display such infrared problems, a feature
that we have exploited elsewhere.\cite{dez}  This emphasizes that this is a
uniquely quantum effect, without a classical counterpart.  One may object that
the Feynman propagator is irrelevant, because of complications defining the
out-state in these coordinates.  But perturbation theory requires the
computation of other interesting quantities such as the time-ordered product
between in-vacuua or, similarly, the anticommutator $G_1$ defined by
\beq
G_1(\tau,\tau',\vec x-\vec x')\equiv \langle 0, in|\{\phi(\tau,\vec x),
\phi(\tau',\vec x')\}|0, in\rangle.
\label{anticomm}
\eeq
These, like the Feynman propagator, suffer infrared divergences.  This
infrared problem is associated with the behavior at large distances and is not
peculiar to large times.\footnote{This divergence is, however, the origin of
the instability discussed in Ref.~\cite{ait}.  Other infrared problems at large
times are the basis for the instabilities discussed in
Refs.~\cite{tw1,tw2,tw4}.}

A useful discussion of this problem has been given by Allen and
Folacci,\cite{af} who concentrate on a different foliation of the de~Sitter
manifold, viz., spatially closed coordinates.  The advantage of these as
opposed to flat (or open) coordinates is that the modes are discrete.  Thus,
the zero mode may be separated and modified, and they introduced modified
vacuua that were suggested by a construction of DeWitt\cite{bsdw} in which the
zero mode is treated differently.  However, such a construction does not
resolve infrared problems for the flat or open coordinates, in which the
momentum is continuous.  Modification of the treatment of the zero mode alone
will not change the cluster properties of the correlation functions.
Consequently, people resort to the introduction of an {\it ad hoc} infrared
cutoff (such
as $|\vec k|>k_0$) whose physical origin is uncertain.  Although the stress
tensor is not infrared divergent, there is little hope that all observables
will be cutoff-independent.  Even for the closed coordinates, it is not so
clear that the so-called Hadamard vacuua discussed in Ref.~\cite{af} are
physically selected.

So the question remains how this infrared problem is resolved
physically.  The existence of a physically reasonable propagator (i.e.,
two-point correlation functions) is a prerequisite to formulating a sensible
perturbation expansion.  Only then can one begin to ask, for example, whether
de~Sitter space is stable to quantum fluctuations.

The outline of the remainder of this paper is as follows:  in the next section,
we discuss the treatment of the nonadiabatic modes.  In Section~3, we evaluate
the change in the stress tensor for our new vacuum state.  In Section~4, we
consider a different method of treating the history, by modifying the metric in
the distant past, and using a WKB expansion for an evaluation of the effects.
In Section~5, we consider yet another modification of the metric, one which
lends itself to analytic solution.  In Section~6, we summarize our results and
comment on their significance.

\section{Nonadiabatic modes}

Even though the stress tensor is known to be free of infrared problems, an
examination of the Hamiltonian does in fact provide some insight into the
nature of the problem.  Because the metric is explicitly time-dependent,
momentum but not energy is conserved.  Nevertheless, even though the
Hamiltonian is not time-independent, it still governs the time-evolution of
the field.  It is useful to construct the Hamiltonian corresponding to the
matter action Eq.~(\ref{matteraction}).  We find
\beq
H(\tau)=\int{d^3k \over (2\pi)^3}k\left[{r_k\over2}\left(
a_{\vec k}^{} a_{\vec k}^{~\dag} + a_{\vec k}^{~\dag} a_{\vec k}^{}\right)
+ {s_k\over2}a_{\vec k}^{}a_{-\vec k}^{}
+ {s_k^*\over2}a_{\vec k}^{~\dag} a_{-\vec k}^{~\dag} \right],
\label{hamiltonian}
\eeq
where
\beq
r_k\equiv \left(1+{ 1\over{2(k\tau)^2} }\right),\hskip0.5in
s_k\equiv -{ {\exp(-2ik\tau)} \over {k\tau} } \left(i+{1\over{2k\tau}}\right).
\label{coeff}
\eeq
The coefficients $r_k, s_k$ depend only on the magnitude $k=|\vec k|.$
Notice that they are in fact functions of the product $k\tau$ only.
{}From this expression, we see that in general, the Hamiltonian causes
transitions increasing or decreasing the number of quanta by two.  In
particular, the Hamiltonian makes transitions between the
de~Sitter invariant vacuum $|0, in\rangle$ defined in Eq.~(\ref{dsvacuum})
and the two-particle states
\beq
|\vec k,-\vec k\rangle\equiv a_{\vec k}^{~\dag} a_{-\vec k}^{~\dag}\
|0,in\rangle.
\label{twoparticle}
\eeq
Note that the this form of the Hamiltonian is precisely that of a BCS
superconductor in quadratic approximation,\footnote{See any modern text on
superconductivity.\cite{bcs}} except that the quanta here are
bosons rather than electrons.  Nevertheless, one is tempted to diagonalize the
Hamiltonian at some particular time by a Bogoliubov
transformation\cite{bcs, nnb} and to try to define no-particle states with
respect to the transformed basis.  This naive expectation is not correct, and
the vacuum state would be extremely sensitive to the time chosen.  Such a
procedure has been a highly disreputable thing to suggest for many
years,\cite{sf, bdbook} especially because it leads to an infinite density of
particles at any later time.  In any case, the off-diagonal $s_k$ vanishes in
the distant past ($\tau\rightarrow-\infty$), so one would be inclined to choose
the in-state at $\tau=-\infty$ to be precisely $|0, in\rangle$, especially
since the rate of change of the metric in conformal coordinates is so slow in
the distant past, and the Hamiltonian approaches a finite, time-independent
limit.

This line of reasoning is perfectly sensible and would be the end of the story
were it not for
the fact that the theory remains infrared divergent in that state.  Moreover,
in our present thinking about cosmology, in which the inflationary phase
arises either after cooling from a hot initial phase or from some sort of phase
transition at very early times, in fact it is
an idealization to presume one is in the de~Sitter phase at all times.  Let us
consider how our description might change as a result of the supposition that
the idealized Lagrangian that we have been discussing is only valid for times
exceeding some early time $\tau_0$.  Our concern is with what the state of the
universe must be.  At first, one would think that the no-particle state will
correspond to the usual adiabatic vacuum.\footnote{The adiabatic vacuum was
developed by Parker\cite{lp} and others and is reviewed in
Ref.~\cite{bdbook}.}  In this construction, at any given time, the modes of
the field may be divided into high- and low-momentum modes, the transition
being set by the time rate of change of the scale factor.  In conformal time,
this is
\beq
{d\over{d\tau}}\log\left(a(\tau)\right)={1\over\tau}.\label{adiabatic}
\eeq
Since we may not entertain times earlier than $\tau_0$, we may think of those
momenta with $k|\tau_0|<1$ as non-adiabatic and those with $k|\tau_0|>1$ as
adiabatic.  For the adiabatic modes, we will assume that the standard
construction prevails, so that, to a good approximation, the correct vacuum
state will be like $|0, in\rangle$ in that it will be annihilated by the
$a_{\vec k}$, but only for $k>1/|\tau_0|.$

However, the
nonadiabatic modes require further discussion.  It is reasonable to ask
whether an observer might design an experiment to probe these low frequency or
very long wavelength modes.  In fact, because of the horizon in de~Sitter
space, they are impossible to detect.  We allude to the well-known fact is
that,
no matter how long one waits, an observer at the origin can only receive
signals that originate later than $\tau_0$ at comoving coordinates $|\vec
x|<|\tau_0|$.  Correspondingly, by the uncertainty principle, a detector could
not be built to probe (conformal) momenta $|\vec k|<1/|\tau_0|$, since no
signal can be communicated over coordinates greater than $|\tau_0|$.
Consider in particular the so-called two-particle states defined in
Eq.~(\ref{twoparticle}).  Normally, we think of them as consisting of two
distinct quanta moving in opposite directions.  However, for frequencies
$k<1/|\tau_0|$, it would be impossible to make such a determination.  In fact,
such nonadiabatic modes are essentially indistinguishable from the no-particle
state $|0, in\rangle$ itself.  Thus, we have a situation in which the
Hamiltonian causes transitions between physically indistinguishable states,
the classic situation discussed by Lee \& Nauenberg.\cite{tdlmn}  They argued
quite generally that one must use techniques of degenerate perturbation theory
if one is to avoid encountering infrared divergences.  In particular, they
showed that one must either sum over degenerate states in forming observables
or {\bf diagonalize the Hamiltonian} in the space of degenerate states.  In
the present context, this suggests that, while the definition of the vacuum
$|0,in\rangle$ given in Eq.~(\ref{dsvacuum}) may be correct for the adiabatic
modes, one should modify the vacuum {\it for the nonadiabatic modes} by
reference to a basis in which the Hamiltonian is diagonal.

In an interacting field theory, diagonalizing the Hamiltonian is a formidable
task, but since we are working here with a quadratic Hamiltonian, it can be
carried out explicitly by means of a Bogoliubov transformation
\beq
b_{\vec k}^{}=\alpha_k a_{\vec k}^{} + \beta_k a_{-\vec k}^{~\dag},
\label{bogoliubov}
\eeq
with coefficients subject to the constraints
\beq
|\alpha_k|^2-|\beta_k|^2=1.\label{constraint}
\eeq
$\alpha_k$ and $\beta_k$ depend only on the frequency
$k=|\vec k|$ and not the direction.\footnote{We will display the coefficients
$\alpha_k, \beta_k$ below.}

This discussion suggests the following improved approximation to the vacuum
state of de~Sitter space.  For an arbitrary Bogoliubov transformation,
Eq.~(\ref{bogoliubov}), one may define a vacuum state $|\tau_0, in\rangle$ by
\beq
b_{\vec k}~|\tau_0, in\rangle=0.\label{vacuum}
\eeq
The state is defined by the prescription one gives for the coefficients
$\alpha_k$ and $\beta_k$.  As suggested earlier, we choose the usual
prescription
$\alpha_k=1,\beta_k=0$ for adiabatic modes $k>1/|\tau_0|.$  Thus, for these
modes, this vacuum is the same as the de~Sitter vacuum.  However, for the
non-adiabatic modes, $k<1/|\tau_0|$, we choose these coefficients such that
they diagonalize the Hamiltonian.  (We shall come back to discuss the
transition between the two regimes later.)  While the physical motivation
based on the Lee-Nauenberg theorem seems quite compelling, it remains to be
shown that this prescription is precisely of the sort described by Ford and
Parker\cite{fp}, i.e., it resolves the infrared problems associated, for
example, with the two-point correlation functions.  In particular, since the
Hamiltonian varies with time, it may not be immediately obvious that
diagonalizing the Hamiltonian at some initial time $\tau_0$ will remove the
infrared divergences for all later times.

As noted earlier, the Hamiltonian is explicitly a function of time, but it is
the Hamiltonian $H(\tau_0)$ at the initial time $\tau_0$ that we wish to
diagonalize.   It is a simple exercise to obtain the form of the Bogoliubov
transformation.  We define an angle $\theta_k$ by the phase of $s_k$, i.e.,
$s_k\equiv |s_k|\exp(-i\theta_k).$  Then we find
\beq
{\beta_k\over\alpha_k}=R_k\exp(-i\theta_k),\ {\rm with\ } R_k\equiv
{r_k\over|s_k|} -\sqrt{\left({r_k\over|s_k|}\right)^2-1}.\label{ratio}
\eeq
A useful form for the phase angle is
\beq
\theta_k=2k\tau_0-\gamma_k+\pi,\ \ {\rm where\ } \tan\gamma_k\equiv2k\tau_0.
\label{theta}
\eeq
Without loss of generality, we may choose the
coefficient $\alpha_k$ to be real,
so that the phase $-\theta_k$ is associated with $\beta_k$.  Then, for the
magnitudes, we find
\beq
\alpha_k={ 1\over{\sqrt{1-R_k^2}} },\hskip0.25in |\beta_k|=
{R_k\over{\sqrt{1-R_k^2} } },\hskip.25in k<1/|\tau_0|.\label{magnitudes}
\eeq
The Hamiltonian then takes the form
\beq
H(\tau_0)= \int{d^3k \over (2\pi)^3}~k\left\{b_{\vec k}^{}, b_{\vec
k}^{\dag}\right\},\label{hdiag}
\eeq
As remarked earlier the state so constructed resembles superconductivity, as
the vacuum state involves pairing of quanta of equal and opposite momenta.
It is important to recognize that the pairing of nonadiabatic modes does not
correspond to condensation of the scalar field.  The symmetry breaking arises
from off-diagonal long-range order.  Thus, in terms of the original quanta,
there are extremely long-range correlations in the vacuum which resolve the
infrared divergences.  Note, however, that unlike ordinary superconductors,
diagonalization does not result in a mass for the elementary excitations.  Of
course, the expectation value of the Hamiltonian $H(\tau_0)$ is lower in the
state $|\tau_0, in\rangle$ than in the original de~Sitter invariant state $|0,
in\rangle$.  So we may think of this as spontaneous breaking of de~Sitter
symmetry, but it also exhibits aspects of explicit breaking, since it involves
a parameter ($\tau_0$) that is not part of the original Lagrangian.  On the
other hand, if we
regard the gravitational action as being defined only on the interval of time
$\tau_0<\tau$, then one may say it was present in the theory from the
beginning.

In the transformed basis, the expansion of the field becomes
\beq
\phi(x)= \int {d^3k \over (2\pi)^3 \sqrt{2k}}\left[
b_{\vec k}^{}u_k (\tau)e^{i{\vec k}\cdot{\vec x}} + b_{\vec
k}^{~\dag}u(\tau)^*e^{-i{\vec k}\cdot{\vec x}}
\right],\label{newfield}
\eeq
where the modified wave functions are
\beq
u_k(\tau)=\alpha_k^{}\phi_k(\tau)-\beta_k^{*}\phi_k^*(\tau_0).
\label{wavefunction}\eeq It is useful to have approximate expressions for
these new wave functions for small momenta.  In the nonadiabatic regime,
$k|\tau_0|\ll1,$ we then find $R_k\approx1-2(k\tau_0)^2$, so that
\begin{eqnarray}
\alpha_k&=&{1\over{2k|\tau_0|}}\left(1+{1\over2}(k\tau_0)^2+\ldots\right)
\nonumber\\
|\beta_k|& = &{1\over{2k|\tau_0|}}\left(1-{3\over2}(k\tau_0)^2+\ldots\right)
\label{nonadiabatic}\\
\theta_k & = &\pi+{4\over3}(k\tau_0)^2+\ldots \nonumber
\end{eqnarray}
In particular, $\alpha_k+\beta_k\rightarrow~O(k|\tau_0|)$ as $k\rightarrow0$,
as required for removable of the infrared singularity.\cite{fp}  To see this
explicitly, we may use these results to obtain the behavior of the
wavefunction,
Eq.~(\ref{wavefunction}), in the nonadiabatic regime.  Since the conformal
time is in the range $\tau_0<\tau<0,$ we may expand for all time to obtain
\beq
u_k(\tau)\approx iH\tau_0\left[1+{1\over2}\left(
(k\tau)^2+(k\tau_0)^2\right) \right]\label{lowk}
\eeq
Thus, the $1/k$ infrared behavior as $k\rightarrow0$ characteristic of the
de~Sitter wave function, Eq.~(\ref{dswave}), has been replaced in the new
basis by a constant $a(\tau_0)^{-1}$, no worse than the flat space or
conformal wave function.  Working in this modified vacuum, the infrared
singularities will be removed, and quantities such as the propagator in our
new vacuum will be well-defined.

For example, the anticommutator $G_1$ defined
with respect to the state $|\tau_0\rangle$ becomes
\beq
G_1(\tau,\tau',\vec x-\vec x')=\int {d^3k \over (2\pi)^3{2k}}
\left[u_k(\tau)u_k^*(\tau')e^{i{\vec k}\cdot({\vec x}-{\vec x'})} +
c.c.\right]. \label{gone}
\eeq
For $k|\tau_0|>1$, the integrand is unchanged from the de~Sitter-invariant
vacuum, but for $k|\tau_0|<1$, the integrand is modified in such as way that
it tends to a constant.  If one considers large spacelike separations, e.g.,
$\tau=\tau'$, $|{\vec x}-{\vec x'}|\longrightarrow\infty$, the behavior will
be dominated by wave numbers $k<1/|{\vec x}-{\vec x'}|$.  For this quantum
amplitude, one is not restricted to the causal region, so one may consider the
regime in which $|{\vec x}-{\vec x'}|\gg |\tau_0|$.  In this regime, the
dominant contribution will come from the nonadiabatic modes where the wave
functions may be approximated by Eq.~(\ref{lowk}).   Thus, for large
separations, $G_1$ will fall to zero,
\beq
G_1(\tau,\tau,\vec x-\vec x')\sim
{{H^2\tau_0^2}\over{({\vec x}-{\vec x'})^2}}.\label{asymp}
\eeq
Thus, this vacuum state satisfies cluster decomposition of the fields.  On the
other hand, the correlator $G_1$ will increase logarithmically out to
separations on the order of $|\tau_0|$ before turning over, so that pairing
only significantly changes the behavior outside the causally connected domain.
Nevertheless, this is important for understanding the behavior of
radiative corrections.   This
damping of long wavelength modes could also have implications for the sort of
vacuum fluctuations that emerge at the completion of an inflationary epic.

At the other extreme, consider the behavior of $G_1$ at short distances.  Of
course, the correlator blows up as in flat space, but the first contribution
to the short distance expansion that depends on the structure of spacetime
comes from $\langle\phi(\tau,\vec x)^2\rangle$.  By translation invariance of
de~Sitter space in conformal coordinates, we may choose $\vec x=0$. Of course,
an operator such as $\phi^2$ requires renormalization in any case.  However,
it has been argued \cite{adl,aas,vf,lf} that, in the usual, de~Sitter vacuum,
the renormalized vacuum expectation value grows as $\ln|\tau|$ owing to the
infrared divergence in this state.  In our new vacuum $|\tau_0\rangle$, this
is not IR divergent.  Thus, any $\log|\tau|$ dependence must then come from
ultraviolet (conformal) momenta, so that it becomes a question of
renormalization.  Since $\phi^2$ is not an observable operator in the present
theory, it is something of a moot point, but we will return to it in
subsequent sections.

One might entertain other states that are equally good candidates as vacuua.
For example, in order to make the transition from the nonadiabatic to adiabatic
modes smooth, one would be tempted simply to adopt the form $\beta_k$ given in
Eqs.~(\ref{ratio}) and (\ref{magnitudes}) for {\it all} $k$, not just the
adiabatic modes, but then to multiply the adiabatic modes by a damping factor,
such as $\exp(-bk|\tau_0|)$ for some constant $b$ of order 1.  Correspondingly,
the magnitude of $\alpha_k$ would be modified so that the constraint
Eq.~(\ref{constraint}) is satisfied.  This seems to us to be a perfectly
acceptable definition of an adiabatic vacuum, although probably more can be
said about the transition between nonadiabatic and adiabatic modes.  These
issues would seem to depend on the details of the history of the de~Sitter
phase.

The usual objections to defining the no-particle state by reference to
diagonalizing the Hamiltonian do not apply to our construction.  Normally, one
would be concerned that there would be an infinite density of particles
produced at all times later than the reference time $\tau_0,$ but that is not
the case here since it is only the very low frequency modes that are
significantly modified.  Moreover, we have argued that one may not observe the
nonadiabatic modes as particles; at best, their effects could be felt
indirectly, for example, through their contribution to the energy density.  In
fact, we will show that their contributions to the energy density are small,
vanishing with $\tau_0$ as $\tau_0^{-2}$.

The preceding considerations are naturally relevant to the vacuum state of
quantum cosmological gravity (QCG), the theory of simply the gravitational
field in the presence of a non-zero cosmological constant but without any
matter.  As mentioned earlier, the small fluctuations about the de~Sitter
background involve gravitons that include modes which, in the linear
approximation, satisfy the same equation of motion as massless,
minimally-coupled scalars.\footnote{This has been reviewed in Refs.~\cite{tw2,
tw3, dez}.}  Thus, we expect the vacuum state of QCG to resemble
$|\tau_0\rangle$, in which the non-adiabatic gravitons of opposite helicity and
momenta are paired.  With some additional effort, such a description can be
made gauge-invariant.  The situation is analogous to a BCS superconductor; the
gap parameter is not gauge-invariant, so pairing would appear to break gauge
invariance.  But it can be shown that the Ward identities are fulfilled, so it
is only the global $U(1)$-symmetry that is broken.  Similarly here, covariant
conservation of the stress tensor is intact, even though global de~Sitter
symmetry is broken.  Unlike the superconductor, however, there are no Goldstone
bosons associated with this symmetry breaking.  In this respect, our situation
resembles explicit breaking:  even though the equations of motion remain
de~Sitter invariant, the action over the interval $\tau>\tau_0$ is not
de~Sitter invariant.

\section{Stress tensor in the state $|\tau_0,in\rangle$}

As another application of this vacuum, consider the renormalized, one-loop
stress tensor $ T_{\mu\nu}$.  Even in the de~Sitter-invariant state
$|0,in\rangle$, it is not beset by infrared divergences, and its vacuum
expectation value is finite, although its value depends on the renormalization
prescription.  To gain insight into the effects of modifying the vacuum in this
way, we evaluate its behavior at one-loop order in the state
$|\tau_0,in\rangle$.  A direct calculation of the {\it unrenormalized}
contribution gives for the nonzero elements
\begin{eqnarray}
\langle T_{00}\rangle&=&{1\over2}\int {d^3k \over (2\pi)^3{2k}}
\left(|{\dot u}_k(\tau)|^2 + k^2 |u_k(\tau)|^2\right)\nonumber\\
\langle T_{ii}\rangle&=&{1\over2}\int {d^3k \over (2\pi)^3{2k}}
\left(|{\dot u}_k(\tau)|^2 - {k^2\over3} |u_k(\tau)|^2\right).
\label{unrenorm}
\end{eqnarray}
We rewrite the various terms in the integrand as, for example,
\beq
|u_k(\tau)|^2= \left(1+2|\beta_k|^2\right) |\phi_k(\tau)|^2
-2\Re{\rm e}\left\{\alpha_k\beta_k\phi_k^2(\tau)\right\},\label{expansion}
\eeq
where we used $|\alpha_k|^2=1+|\beta_k|^2$ in the first term.  The value when
$\beta_k=0$ is of course the de~Sitter invariant contribution, whereas the
terms involving $\beta_k$ represent the change arising from our modification of
the vacuum.  The de~Sitter invariant contribution is completely independent of
$\tau_0$.  It gives
rise to quartic and quadratic ultraviolet divergences which, in a
mass-independent renormalization scheme such as minimal subtraction, is
completely removed by counterterms.\footnote{People often adopt a
renormalization scheme that retains a finite value for this piece, but it is
arbitrary.}  The process of renormalization also introduces a well-known
anomalous contribution\cite{duff} to the stress-tensor which will be small for
$H^2/\mpl^2$ small.  The remaining terms involving $\beta_k$
depend on the initial time $\tau_0$ and cannot be removed by renormalization.
As our interest is in the difference between the stress tensor in the de~Sitter
vacuum and the stress tensor our new vacuum $|\tau_0,in\rangle$, we will ignore
the anomalous contribution here.  Thus, the change in the contributions to the
one-loop renormalized stress tensor is
\beq
\langle T_{00}\rangle={1\over{2\pi^2}}\int_0^{1/|\tau_0|} {kdk}\left\{
\left[ |\beta_k{\dot\phi}_k|^2 -\Re{\rm e}
\left(\alpha_k\beta_k{\dot\phi}_k^2\right)
\right]
+ k^2\left[|\beta_k\phi_k|^2 -\Re{\rm e}\left(\alpha_k\beta_k\phi_k^2\right)
\right]\right\},\label{renorm}
\eeq
with a similar expression for $\langle T_{ii}\rangle$.  Although the
expressions for $\beta_k$ are quite complicated, we may use simple scaling
arguments to obtain the general form of $\langle T_{\mu\nu}\rangle$.  First of
all, note that $|k\phi_k|^2=H^2(1+(k\tau)^2)$ and
$|{\dot\phi}_k|^2=H^2(k\tau)^2$.  Since $\alpha_k$ and $\beta_k$ are functions
only of $k\tau_0$, we may rescale $k\rightarrow k/\tau_0$ to see that $ \langle
T_{00}\rangle$ is of the form $A/\tau_0^2+B\tau^2/\tau_0^4,$ where A and B are
numerical constants that depend on the detailed form of the function $\beta_k$.
Therefore, the energy density and pressure are
\begin{eqnarray}
\langle T_0^0\rangle&\equiv\rho=&
H^4\left[+A~{\tau^2\over\tau_0^2}+B~{\tau^4\over\tau_0^4}\right]\label{energy},
\\ -\langle T_i^i\rangle&\equiv p=&
H^4\left[-{A\over3}~{\tau^2\over\tau_0^2}+{B\over3}
{}~{\tau^4\over\tau_0^4}\right],{}~~({\rm not~summed~on~}i.)\label{pressure}
\end{eqnarray}
It is easy to check that these forms obey the covariant conservation
constraint $T_{\mu\nu}^{~;\nu}=0,$ providing an independent check on the
relative signs appearing in the expression above. While our modified vacuum
state contributes nonzero energy
density (with $\langle T_{\mu\nu} \rangle$ not proportional to
the metric tensor $g_{\mu\nu}$),
these effects rapidly die away.
(Recall that $-H\tau=\exp(-Ht)$, so they decay exponentially in time.)  The
signs of the pressure provide insight into the two terms.  The $B$-term is
traceless and has the form of a relativistic gas that is diluted as
$a(\tau)^{-4}$.  The $A$-term has negative pressure and falls only as
$a(\tau)^{-2}.$  These are the amplified waves originally discussed by
Grishchuk.\cite{lpg}  Finally, as expected, as $\tau_0\rightarrow-\infty$,
these contributions vanish.

Of course, because $\langle T_{\mu\nu}\rangle$ is non-zero, there will be
feedback on the background metric, so the de~Sitter form is no longer exact.
However, for $H^2/\mpl^2\ll1,$ this is a small distortion which, according to
Eqs.~(\ref{energy}) and (\ref{pressure}), decays with time.

\section{Modifications of the metric as an infrared cutoff.}

In this section we will pursue the same idea of introducing an infrared cutoff
through the finite time of existence of de~Sitter phase but in a
different manner. Namely we will make an ad hoc assumption that metric has form
(4) with the scale factor $a(\tau )$ equal to
\beq
a^2(\tau)~=~{1\over {H^2\tau^2}}+{{\delta}\over{H^2}} \label{modified}
\eeq
so that at large times, or $\tau \rightarrow 0$ we
recover de~Sitter space. On the other hand assuming $\delta$ to be
positive we see that  as $\tau \rightarrow -\infty$, or in the distant past,
the metric becomes that of Minkowski spacetime. To be consistent with the
Einstein's equations we tacitly assume a nontrivial matter distribution which
would drive to the solution (\ref{modified}) (see also below).
We will demonstrate that finite $\delta$ does induce an infrared cut
off both when one evaluates $\langle 0|\phi ^2|0\rangle$ and $\langle 0|
T_{\mu\nu}|0\rangle$. What might be most amusing, is that the
infrared cutoff manifests itself in these two cases in different fashion.

Since the metric has been modified we need to find the corresponding modes
of the scalar field which would replace now solutions (\ref{dswave}).
We can try either to solve the equations of motion for the scalar field
or to apply WKB expansion.
The former approach has a shortcoming of being somewhat cumbersome while
the latter, at first sight, is not suited at all to study infrared problems
we are interested in. In fact the both ways work well. In this section
we apply the
high-frequency, or WKB expansion, while discussion of the exact solution is
postponed until the next section.

Let us first explain the general strategy of dealing with infrared problems
within the WKB expansion and introduce to this end the necessary notation.
We are looking for a solution $\phi_{{\bf k}}$
with 3-momentum ${\bf k}$ which is represented as
\beq
\phi_{{\bf k}}({\bf x})~=~{{\exp(i{\bf k\cdot x})}
\over{(2\pi)^{3/2}a(\tau)}}\chi_ k
(\tau)\eeq
where $\chi_ k$ in massless case satisfies the
equation (see, e.g., \cite{bunch,bunch1}):
\beq
{{d^2\chi_k}\over {d\tau^2}}+\{k^2-(\xi-{1\over 6})
a^2R\}\chi_k~=~0
\label{qc}.\eeq
Here $R$ is the Ricci scalar which can be computed as
\beq
R~=~a^{-2}\left(3\dot{D}+{3\over 2}D^2\right),~~D~=~2\dot{a}/a
\eeq
and the dot denotes the derivative with respect to the conformal time $\tau$.
The quasiclassical solution to (\ref{qc}) is
\beq
\chi_k~=~{{\exp(-i\int W_{{\bf k}}(\eta')d\eta')}
\over {(2W_{{\bf k}}(\tau ))^{1/2}}},\eeq
 \beq W_k^2~=~k^2+(\xi -{1\over 6})a^2R-
{{\stackrel{..}{W_k}}\over {2W_k}}+{3\over 4}\left({{\dot{W_k}}\over {W_k}}
\right)^2\label{qcl}\eeq
Eq.~(\ref{qcl}) allows to obtain $W_k$ as an expansion
in $k^{-2}$ in a regular way.

In particular, the first two terms in the expansion are well known:
\beq
W_k^2~=~k^2\left(1-{{(\xi -1/6)a^2R}\over {2k^2}}+O(k^{-4})\right) \eeq
or
\beq
\phi_{{\bf k}}~=~(2\pi)^{-3/2}\exp(i{\bf k\cdot x}-ik\tau)\left(
\tau+i{{(\xi -1/6)\tau a^2R}\over {2k}}...\right)\label{two}
\eeq
while further terms can be obtained iteratively.

What is special about de~Sitter background and $\xi = 0,1/6$
is that these first two terms
(\ref{two}) reproduce the exact result (see Eq.~(\ref{dswave})). It is worth
emphasizing that
this termination of the expansion can be established within the quasiclassical
expansion itself. The coefficient in front of $k^{-3}$ term in the expansion
(\ref{two}) does turn out to be zero if $Ra^2=12/\tau^2$.

As a rule evaluation of a physical quantity involves integration over modes.
Consider for example $\langle0|\phi^2|0\rangle$. From Eq.~(\ref{two})
one immediately gets \cite{bunch}
\beq
\langle 0|\phi^2|0\rangle ~=~{{1}\over {(2\pi)^3}}\int {{d^3k}
\over{2 k}}\left(
1-{{(\xi -1/6)a^2R}\over {2 k^2}}+O(k^{-4})\right)\label{vev}
.\eeq
We will return later to the question of the ultraviolet regularization.
In fact, the integrand is obtained in the high momentum limit, and the exact
one should be infrared finite.  However, the signal of the infrared cutoff may
be inferred from the higher-order terms like $k^{-4}.$
In case that one starts with a
quantity
which is ultraviolet divergent (like $\langle0|\phi^2|0\rangle$), the following
two simple rules seem to apply. First, all power-like ultraviolet divergencies
can be removed through subtractions so that we have to concentrate on the log
term. Clearly, the coefficient in front of the log is uniquely fixed by the
$k^{-2}$ expansion. Second, the next term in the expansion after the
logarithmic one, i.e. the first term which diverges as a power in the infrared
provides with an estimate of the infrared cutoff in the log.  Indeed the
infrared cutoff can be estimated by finding at which $k^2$ the term producing
the log and the next one are of the same order.

{}From this point of view, infrared problems in the de~Sitter background arise
not because subsequent terms in the $k^{-2}$ expansion (see Eq.~(\ref{two}))
produce power-like infrared divergences but because for $\xi=1/6$ the
high-frequency expansion terminates, and there are in fact no further terms in
(\ref{vev}). As a result there is no infrared cutoff for the log and
$\langle0|\phi^2|0\rangle$ is undetermined. This is just the conclusion which
was drawn from analyzing the exact solution for $\phi_k$.
Now we expect that by introducing $\delta \neq 0$ (see Eq.~(\ref{modified})) we
destroy the conspiracy of the coefficients which terminates the expansion
(\ref{two}) and there emerges an infrared cutoff. We proceed now to check these
expectations explicitly.
To first order in $\delta$
\beq
Ra^2=~{12 \over {\tau^2}}(1-{1\over 2}\delta \tau^2)
.\eeq
Let us note that as far as Eq.~(\ref{qcl}) is concerned such a modification of
the metric is equivalent perturbatively to introducing positive $\xi\neq 0$,
\beq
\xi_{eff}~\approx~3\delta\tau^2
.\eeq
As is well known,\cite{bdbook}  there is an important difference between
positive and negative $\xi$. Namely, negative $\xi R$ is like a negative
mass-squared inducing an instability that leads to spontaneous breakdown of the
 de~Sitter-invariant vacuum state.  As a result, one cannot match the WKB
expansion with the correct infrared behavior.  If $\xi $ is positive, there is
no danger of mismatch between the infrared and ultraviolet expansions.

Keeping terms linear in $\delta$ and expanding $\phi_k$ in $k^{-1}$
leads to the following results:
\beq
{{\phi_{{\bf k}}}\over H}={{e^{(i{\bf k\cdot x}-ik\tau)}}\over {(2\pi)^3}}
\left(\tau-{i\over { k}}+0{1\over {\tau k^2}}
+0 {i\over {\tau^2 k^3}}+0 {1\over{\tau^3 k^4}}
-{{i\delta \tau}\over {2k}}-{{\delta\tau}\over {k^2}}+{{i 3\delta}\over{4k^3}}
+0{{\delta}\over{\tau k^4}}+O(k^{-5})\right)\label{expan}
.\eeq
Here we indicated the vanishing coefficients as well since they illustrate
the statement above that the WKB expansion is perfectly consistent with the
exact result for $\phi_k$ (at $\delta =0$).

Now we are in position to apply the rules of defining the infrared cutoff
mentioned above. Indeed, we see that the series in $k^{-2}$ which represents
$\langle 0|\phi^2 |0\rangle$ does have further terms. The infrared cutoff,
$k_{IR}$ can be estimated by equating two subsequent terms in the expansion:
\beq
k_{IR}^{-1}~\sim~\delta k_{IR}^{-3},~~~k^2_{IR}~\sim~\delta \label{irc}
.\eeq
Note that the infrared cutoff (\ref{irc}) is time independent.  Thus, for
$\langle 0|\phi^2|0\rangle$ we find (omitting the subtraction term associated
with the UV quadratic divergence)
\beq
\langle 0|\phi^2|0\rangle~=~{H^2\over {4\pi^2}}\ln \Lambda^2_{UV}
/\delta,~~\xi=1/6~
.\eeq
According to the standard arguments the ultraviolet cutoff, $\Lambda_{UV}$
does depend on time in the conformal coordinates,
\beq
\Lambda_{UV}~\sim~M a(\tau)\label{uv}
\eeq
where $M$ is some mass parameter. Indeed, all the masses are scaled with
$a(\tau)$ in de~Sitter background.
If so, we reproduce the well-known result\cite{adl,aas,vf} for $\langle 0|
\phi^2|0 \rangle$:
\beq
\langle 0|\phi^2|0\rangle~\sim~-{{H^2}\over {2\pi^2}}\ln\tau
.\eeq

We turn now to consideration of the vacuum expectation value $\langle 0|
T_{\mu\nu} |0\rangle$. The change brought by $\delta \neq 0$ is a
multiplicative factor $(1-3\delta/2k^2)$ as far as energy and momentum
associated with each mode $\phi_{{\bf k}}$ are concerned. If we turn to the
integral over modes determining the vacuum expectation value then this factor
is crucial since the term proportional to $\delta$ gives a logarithmically
divergent integral over $k$. Keeping this log term we find
\beq
\langle 0|T_{00}|0\rangle~=~-3\langle 0|T_{ii}|0 \rangle~=~-
\delta{3H^2\over {8\pi^2}} \cdot \ln{{\Lambda_{UV}}\over
{\Lambda_{IR}}}\label{log}\eeq
where $\Lambda_{UV},\Lambda_{IR}$ are ultraviolet and infrared cutoffs and no
summation over $i$ is assumed.  Moreover, as explained above, we do not keep
power-like divergent terms.

For the ultraviolet cutoff we still have (\ref{uv}). It is most interesting
that the estimate (\ref{irc}) for the infrared cutoff is now changed. The
reason is that the terms kept in the expansion (\ref{expan}) are sufficient
only to find the coefficient in front of the log term (\ref{log}) while for the
estimate of the infrared cutoff we need the next term in the $k^{-2}$ expansion
of the solutions $\phi_{{\bf k}}$.  One can obtain the following estimate
\beq
\Lambda^2_{IR}\tau^2~\sim~1\label{irrr}
.\eeq
It might worth noting that the time dependence exhibited by (\ref{irrr}) is so
to say normal, since the WKB expansion is an expansion in inverse powers of
$k\tau$. The fact that this time dependence did not manifest itself in
(\ref{irc}) is quite unique and is due to the termination of the WKB expansion
in pure de~Sitter background.

Thus, we have for the contribution to the vacuum expectation value of the
stress tensor:
\begin{eqnarray}
\rho~\equiv~\langle 0|T^0_0|0 \rangle \propto\cdot H^4
\tau^2\delta,&~& p~=~-{1\over 3}\rho\label{new}.
\end{eqnarray}
Two remarks concerning this result are now in order. First, the vacuum
expectation value is to satisfy the covariant conservation condition
\begin{eqnarray}
D_{\mu}\langle 0|T^{\mu}_{\nu}|0\rangle~=~0,& {}~{\rm or}~&\tau{d\over
{d\tau}}{\rho}~=~3(\rho+p)\label{cons}
\end{eqnarray}
where the latter equation makes use of de~Sitter metric, and the energy density
$\rho$ and pressure $p$ are defined in Eqs.~({\ref{energy}, \ref{pressure}).
The use of de~Sitter metric is justified since $\langle 0|T_{\mu\nu}|0 \rangle$
starts from terms proportional to $\delta$ and in the linear approximation one
can neglect that $\delta\neq 0$ otherwise.

The condition (\ref{cons}) establishes a relation between time dependence and
the coefficient of proportionality between $\rho$ and $p$, in our case
$p=-1/3\rho$.  Moreover it is worth emphasizing that constraint (\ref{cons})
should be satisfied both for each mode with fixed ${\bf k}$ and for the sum
over the modes, or the vacuum expectation value. Since for a logarithmically
divergent integral the relation between $\rho$ and $p$ for the sum over modes
is the same as for each mode one immediately concludes that no extra time
dependence can be associated with the $\ln(\Lambda_{UV}/\Lambda_{IR})$. This is
a nontrivial check of general covariance which is passed by the estimates
above.

Second, Eq.~(\ref{new}) implies that $\langle 0|T_{\mu\nu}|0 \rangle$ is not
proportional to $g_{\mu\nu}$ which in de~Sitter background would result in
$\langle 0|T_{00}|0 \rangle =- \langle 0|T_{ii}|0\rangle $. If we were
considering pure de~Sitter solution (i.e. $\delta =0$) then $g_{\mu\nu}$ would
be the only tensor structure available, and Eq.~(\ref{new}) would not be
possible. However, in the approximation linear in $\delta$, we do have a new
tensor structure emerging in $R_{\mu\nu}$:
\beq
R_{00}~=~{3\over {\tau^2}}(1+\delta \tau^2),~~R_{ii}~=~-{3\over {\tau^2}}
(1-\delta \tau^2).
\eeq
By virtue of equations for the gravitational field the term
proportional to $\delta$ reflects the structure of $T_{\mu\nu}$ of matter
needed to produce solution (\ref{modified}). This new tensor
structure is also manifested in Eq.~(\ref{new}).

Thus, by assuming the scale factor to be given by (\ref{modified})
we have tacitly introduced nontrivial stress energy of matter
which drives metric to such a solution. Then contribution
(\ref{new}) can be viewed as a radiative correction to the stress tensor
of the matter. The plausible interpretation of this
correction is that during the transition from Minkowski to de~Sitter
phases scalar particles are produced.

In conclusion of this section, let us compare this method with that of the
preceding sections involving the introduction of the initial time $\tau_0.$
Note that, although technically the prescriptions for the infrared cutoff look
very similar since they modify the situation in the distant past, the physics
is actually quite different in these two cases. Indeed, in the latter case we
assume that there is some matter distribution which causes
$\langle0|T_{\mu\nu}|0\rangle\neq const\cdot g_{\mu\nu}$. In the former case,
the deviations from pure de~Sitter solution are entirely due to the initial
state.

\section {Exact solution for a modified metric.}

Here we will illustrate some of the results discussed above for the particular
case of a modified de~Sitter metric which was originally flat
(at negative time infinity) and which admits an exact solution for the
minimally coupled scalar field $\phi$. We chose the metric to be of the
usual Robertson-Walker form with the scale factor:
\beq{
a(t) = 1 + \exp (Ht) = [1-\exp (H \tau )]^{-1}
}\label{aoft}
\eeq
where the conformal time $\tau$ is expressed through $t$ as
\beq{
\tau = -H^{-1} \ln [1 +\exp (-Ht) ]
}\label{conft}
\eeq
As above the future infinity, $t\rightarrow +\infty$,
corresponds to $\tau\rightarrow-0$.
The scale factor (\ref{aoft}) cannot be created by any realistic source.
For large positive time the source is close to the cosmological term and
this can be achieved but in the past when both the energy and pressure
density were close to zero the pressure should be negative and much larger
by magnitude than energy. The latter cannot be realized by normal physical
matter. Still this metric is convenient for the analysis of the quantum
fluctuations in the de~Sitter spacetime because the
equation of motion (\ref{eom}) can be exactly solved. The solution
which behaves as $\exp (ik\tau)$ when
$\tau \rightarrow -\infty$, has
the form:
\beq{
\phi_k = y^{i \kappa }
F(\alpha, \beta, \gamma; y)
}\label{hyper}
\eeq
where $\kappa =k/H$,
$F(\alpha, \beta, \gamma; y)$ is the hypergeometric function as determined e.g.
in ref. \cite{grry},
$\alpha =  -1 + \sqrt{1-\kappa^2} +i\kappa$,
$\beta =  -1 - \sqrt{1-\kappa^2} +i\kappa$,
$\gamma=1+2i\kappa$, and $y=\exp(H\tau)$. When $\tau$ changes from $-\infty$ to
$-0$, $y$ changes from 0 to 1.

In the infrared limit when $k\rightarrow 0$ and $y$ is fixed,
the solution is well defined,
\beq{
F\rightarrow 1 + i\kappa (-2y + y^2)
}\label{fred}
\eeq
in contrast to expression (\ref{dswave}) which diverges as $1/k$.
One would expect however that this divergence is recovered after a long
de~Sitter stage when $y\rightarrow 1$ (or $\tau \rightarrow -0$). This
particular hypergeometric function may be approximated near $y=1$ as
\beq{
F\approx {\Gamma (\gamma)\over \Gamma (\gamma - \alpha)
\Gamma (\gamma - \beta)}
\left[2 + \gamma (y-1) + {\gamma (\gamma +1)\over 2} (y-1)^2 \right]
}\label{fy1}
\eeq
Once again the limit of small $k$ is nonsingular and agrees
with expression (\ref{fred}) for $y=1$.
We will observe
the trace of the above mentioned infrared singularity in the ultraviolet
expansion, $\kappa \rightarrow \infty$, of the solution. The result depends
upon the way the limits $k\rightarrow \infty$ and $y\rightarrow 1$ are taken.
For the case of $y\rightarrow 1$ first and then $k\rightarrow \infty$,
expression (\ref{fy1}) is convenient. We find from it (up to
$y$-independent phase factor) that
\beq{
F\approx \left(1- {\pi\over 4\kappa} +
{3+2\pi^2 \over 24\kappa^2}  \right)
\left[ {1\over \kappa} + \left( i +
{1\over 2\kappa} \right) (y-1) - \kappa \left(1-{i\over 2\kappa}\right)
\left(1-{i\over \kappa} \right) (y-1)^2 \right]
}\label{f1k}
\eeq
The previously found solution for the exact infinitely long de~Sitter stage
is reproduced by the first two terms in the expansion in $(y-1)$ in the
limit of \underline{large} $k$. The infrared singularity at $k=0$
in the exact de~Sitter solution (\ref{dswave}) is evidently absent in the
distorted metric and appears here only in the ultraviolet expansion.
Note that due to the factor in front of the r.h.s. of Eq.~(\ref{f1k}) the
expansion of $|F^2|$ contains not only even but also odd powers of $1/\kappa$.
It would generate new types of divergences in perturbation theory.

All these phenomena may be related to a singular character of the limiting
transition from the metric (\ref{aoft}) to the exact de~Sitter one. Indeed let
us introduce a small parameter $\epsilon$ (instead of 1 in Eq.~(\ref{aoft}))
which characterizes a deviation from the de~Sitter metric at a fixed finite
time, so the metric has the form:
\beq{
\tilde a (t) = \epsilon + \exp (Ht)
}\label{atil}
\eeq
By shifting the zero of time $t=t'-t_1$ so that $\exp (-Ht_1) =\epsilon$,
we may rewrite $\tilde a(t)$ as $\tilde a(t')=\epsilon [1+\exp(Ht')]$.
Now by rescaling the space coordinates $dx=\epsilon dx'$ we return to
the old metric (\ref{aoft}). These transformations correspond to rescaling
of the conformal momentum $k\rightarrow k/\epsilon$. For finite $\epsilon$
this change of normalization is unobservable. But the de~Sitter limit,
when $\epsilon\rightarrow 0$, is singular and gives rise to the peculiar
behavior mentioned above.

It may be also interesting to get ultraviolet expansion of the solution
(\ref{hyper}) for a fixed $y$.
To get this
expansion it is convenient to present $F$ in the form
\beq{
F= 1-y + {i\over \kappa } f + {1\over \kappa^2} g
}\label{fult}
\eeq
The functions $f$ and $g$ can be found as expansions in inverse powers of
$\kappa^2$: $f=f_1+f_2/\kappa^2 + f_3/\kappa^4...$ (and the similar expression
for $g$) by the perturbative (in $1/\kappa^2$) solution
of the equations:
\beq{
(1-y)f' + f =-{1+y\over 2} + {1+y\over 2\kappa^2}g' +
{y(1-y)\over 2\kappa^2} g''
}\label{fprim}
\eeq
\beq{
(1-y)g' + g = - {1+y \over 2} f' - {y(1-y)\over 2} f''
}\label{gprim}
\eeq
The terms which are logarithmically divergent in momentum
are determined by the
lowest order corrections $f_1$ and $g_1$. The latter are
singular, though not too strongly, near $y=1$ but this divergence is cancelled
out, as we see in what follows, in the expression for $\langle \phi^2\rangle$.
The calculation is straightforward and we get:
\beq{
f_1 = -y-{1-y\over 2}\ln (1-y)
}\label{f1}
\eeq
\beq{
g_1= {y\over 4} -{y\over 2} \ln (1-y) -{1\over 8} (1-y)\ln^2 (1-y)
}\label{g1}
\eeq
The singularity in $(1-y)$ is getting stronger with the rising order in
$1/\kappa^2$. For example the most singular parts for $f_2$ and
$g_2$ are respectively $f_2^{sing} = 3/8(1-y)$ and $g_2^{sing} =
-1/4(1-y)^2$.

Using these expressions we may calculate the vacuum expectation value of
$\phi^2$ as a function of time. For a pure de~Sitter universe, this quantity
is known to rise as $\ln \tau =t$ \cite{adl,aas,vf}. This result is ascribed
to the infrared instability of a minimally coupled massless scalar field.
We will calculate the difference
$\langle\delta \phi^2 \rangle \equiv \langle \phi^2 (\tau)\rangle
-\langle \phi^2 (-\infty)\rangle$.
One may hope that this automatically takes care of infinities in
flat spacetime. In the limit of small $y$ one may write keeping only
terms of order $y^2$:
\beq{
\langle \phi^2(y) \rangle =\int {d^3k \over (2\pi)^3 2k} \left[ (1-y)^2 +
{2y\over 4\kappa^2 +1 } +
{(2\kappa^2 - 1) y^2 \over (4\kappa^2+1)(\kappa^2+1)} \right]
}\label{vacphi}
\eeq
where $y =\exp (H\tau)$. One has to subtract from this quantity the value
at $y=0$. The first term is quadratically divergent and to specify its value
we have to define the regularization procedure. Usually it is just an arbitrary
renormalization constant but in this case it becomes a function of time
(or $y$). To avoid this time dependence one may change the variable of
integration, $\kappa = a(\tau) p$ which means transition from the conformal
momentum to the physical one. Written in this form
the quadratically  divergent integral in the expression for $\phi^2(y)$
is formally time independent and is cancelled out by the initial value in
the flat spacetime.

There are also logarithmically divergent terms which are absent in the flat
case. For small $y$ they can be found from Eq.~(\ref{vacphi}) which is exact
with respect to the dependence on $\kappa$ and correspondingly is not singular
at $\kappa=0$. Calculating the integral we find
\beq{
\langle \delta\phi^2 \rangle =
{H^2 \over 16\pi^2} y(1+y) \ln \kappa_{max}^2 }\label{delta}
\eeq
Here $\kappa_{max}$ is (unspecified) ultraviolet cutoff. For arbitrary $y$ we
may use Eqs.~(\ref{fult},\ref{f1},\ref{g1}) giving expansion in inverse powers
of $\kappa$. Though the result is formally singular at $\kappa=0$ we know that
the effective infrared cutoff is $\kappa \approx 1$. The logarithmic dependence
on $(1-y)$ which we see in $f_1$ and $g_1$ is cancelled out and we get exactly
the same result (\ref{delta}). The logarithmically divergent terms are
explicitly $y$-dependent and this dependence cannot be cancelled by a
redefinition of the ultraviolet cutoff as it was done above for the
quadratically divergent part.  If we assume that there is a physical cutoff
realized, say, by some massive particles, we would expect the cutoff in
conformal momentum to be proportional to the scale factor   $k=ma(\tau)$. This
is because mass enters equations of motion in conformal coordinates always in
the combination $ma(t)$.  In this case we get the result $\langle \delta \phi^2
\rangle \sim y(1+y)\ln(1-y)$.  At the de~Sitter stage when $y\rightarrow 1$, we
get the well-known result  $\langle \delta \phi^2 \rangle \sim t$ plus a
logarithmically divergent term. Note that the coefficient in front of the log
is exactly $R=6y(1+y)H^2$.  Expressed in terms of $R$ the combination $(1-y)$
looks rather unusual: $(1-y)=\sqrt{1+2R/3H^2} -3$.

Let us turn now to the calculation of the energy-momentum tensor of the field
$\phi$ in background (1). Though it is a more UV divergent quantity than
$\phi^2$, it has the nice property of covariant conservation which helps
greatly in a proper definition of the divergent parts.  Let us start from a
more simple
example when the background metric is flat in the infinite past and future and
changes as a function of time in between.  Assume that the scale factor
$a(-\infty) =1$ and $a(+\infty)=a_f$.  The mode decomposition of $\phi$ has the
standard form (8) with $\phi_k(\tau) = \exp(-i\omega \tau)$ at negative time
infinity and $\phi_k(\tau) = [\alpha_k \exp(-i\omega \tau) + \beta_k
\exp(i\omega \tau)\a_f$ at positive time infinity. The Bogoliubov coefficients
$\alpha_k$ and $\beta_k$ satisfy the relation Eq.~(\ref{constraint}).  We will
use the Heisenberg representation and will calculate the variation with time of
the vacuum energy density. One has in conformal coordinates
\beq{
T_{\tau\tau} ={1\over 2} [(\partial_\tau \phi)^2 + (\partial_j \phi)^2]
}\label{ttau}
\eeq
The physical energy density $\rho$ is expressed through it as $\rho \equiv
T_{tt} =T_{\tau\tau} /a^2$. We will consider the expectation value $\langle
\rho (t) \rangle$ averaged over the initial vacuum state. It is an infinite
quartically divergent quantity, but the difference $\delta\rho = \langle
\rho(+\infty)-\rho(-\infty) \rangle $ is finite and is equal to
\beq{
\delta \rho
\equiv \langle \rho(+\infty ) - \rho (-\infty) ={1\over a_f^4}\int {
2|\beta_k|^2\omega^2\over 2 \omega (2\pi)^3 } d^3k }\label{delrho}
\eeq
It is the energy density of the particles produced by the time varying
gravitational field. In this description we do not need a precise meaning of
``particle", the only essential quantity is the energy density. One can easily
see that the pressure density of the final state is $\delta p= \delta \rho/3$.
We use this simple example to illustrate that the finite part of the
energy-momentum tensor averaged over the in-vacuum state is not proportional to
the metric or to any other tensor associated with the problem and for the
finite parts nonlocal effects are essential.  Note also that the finite time
difference $\delta\rho (t) =\langle \rho(t)-
\rho (-\infty)$ is quadratically divergent, and one needs curvature-dependent
counterterms to eliminate these infinities.

The situation is more complicated for the metric (\ref{aoft}) when the
spacetime is not flat in the infinite future.  Still we may use the similar
approach to calculate the evolution of the energy density of the field $\phi$
starting with the initial vacuum value which we assume to be zero.  The average
value of the energy density over the initial vacuum state is
\beq{
\rho= {1\over a^2(t)} \int { d^3k \over (2\pi)^3}{\omega\over 2} \left[ |F^2| +
{Hy\over k} \Im {\rm m} (F^* F') + {H^2y^2\over k^2} |F'|^2 \right]
}\label{rho1}
\eeq
Here $\rho=T_{tt}$ is the energy density in the physical frame.  Substituting
expressions (\ref{fult},\ref{f1},\ref{g1}) into this equation we get
\beq{
\rho = {1\over a^2} \int { d^3k \over (2\pi)^3}{\omega\over 2}
\left[ (1-y)^2 +{H^2y^2\over k^2} + {H^4\over 2k^4} K(y)
\right]
}\label{rho2}
\eeq
The first two terms, quartically and quadratically divergent respectively, may
be eliminated by renormalization (see discussion in Sec.~3).  The renormalized
quantities are at worst constants so their contribution into $\rho$ is not
rising with time.  The last term is logarithmically divergent, but nonsingular
at $y=1$. The function $K(y)$ might have in principle terms behaving like
$1/(1-y)$ or $\ln^2 (1-y)$, etc., as one can see from Eqs.~(\ref{f1},\ref{g1})
but all they have cancelled out. Hence the energy density of a massless
minimally coupled scalar field in de~Sitter background does not rise with time
but goes down at least as $1/a^2$ and its back reaction on metric is not
essential. Higher loop contributions probably will not change this conclusion,
but this remains an open question.

\section{Conclusions.}

In this paper, we have provided a framework for resolving the infrared problems
associated with the vacuum of de~Sitter space.  As previous workers have
remarked, there is no solution without breaking de~Sitter invariance.  We have
shown that the cosmologically well-motivated assumption that the de~Sitter
phase did not exist indefinitely is sufficient to remove these infrared
divergences.  The detailed modifications will of course depend on the
particular history that precedes the era in which the vacuum energy dominates,
but we have shown that the infrared problem can be resolved without reference
to these details.

We have offered several approaches that may be useful in different contexts.
In the first, we suggested a treatment of the nonadiabatic modes that differs
from the conventional one.  The single new parameter is the initial time
$\tau_0$ before which the vacuum energy is presumed not to be dominant.  In a
manner analogous to the classic Lee-Nauenberg discussion of mass
singularities,\cite{tdlmn} we argued that one must diagonalize the Hamiltonian
before embarking on a perturbation expansion.  We found this to be an
economical formalism, in which the correlation functions differ significantly
only at distances large compared to the distance that a signal might have
traveled since the initial time $\tau_0$.  We showed that the correlation
functions satisfy cluster decomposition, so that this is an acceptable
framework for discussing quantum corrections.

In the second approach, we explicitly modified the metric at early times, in
such a way that, in the distant past, the space is Minkowski spacetime, and we
explored the consequences at later times.  While this cannot be solved
analytically, a WKB solution provided insight into the nature of the infrared
cutoff.  In the third approach, we simply modified the scale factor in a simple
manner, again so the past is Minkowskian, but in a way that the equations of
motion can be solved analytically.

Each approach has its strengths and weaknesses, but the first involving pairing
has the advantage that it presumes little about the universe prior to the
de~Sitter phase, and the dynamical equations remain unchanged, so that their
symmetry properties remain exact.  Whether the transition between nonadiabatic
and adiabatic modes can be refined in some general way, for example, through a
path integral approach,\footnote{See, e.g., Ref.~\cite{dmcjbh}.} is not clear.
The pairing mechanism, as in superconductors, when applied to gravitons may
have far-reaching consequences that we can not now foresee.  Whether, in the
case of gravity, this construction requires modification to take into account
the self-interactions remains to be investigated.  The approach in which the
metric was explicitly modified may be useful when we know the universe is
approximately Minkowski spacetime in the distant past.  It also
illustrates how the WKB method may provide insight into the nature of the IR
cutoff in situations where the nonadiabatic modes cannot be so simply treated.

We emphasize that the considerations of this paper do not resolve the issue of
whether there might be long-time singularities that destabilize the de~Sitter
metric.\cite{tw1,tw2,tw4}  It remains to be seen what consequences, if any, our
results have on interacting field theories such as those considered by
Ford\cite{lf} in which the  behavior of $\phi^2$ directly influences the stress
tensor.
\medskip

\underline{Acknowledgements.}  A. Dolgov is grateful to the
Department of Physics of the University of Michigan for the hospitality during
the period when this work was done.  This work was supported in part by the
U.S. Department of Energy.
\vfill\eject

\end{document}